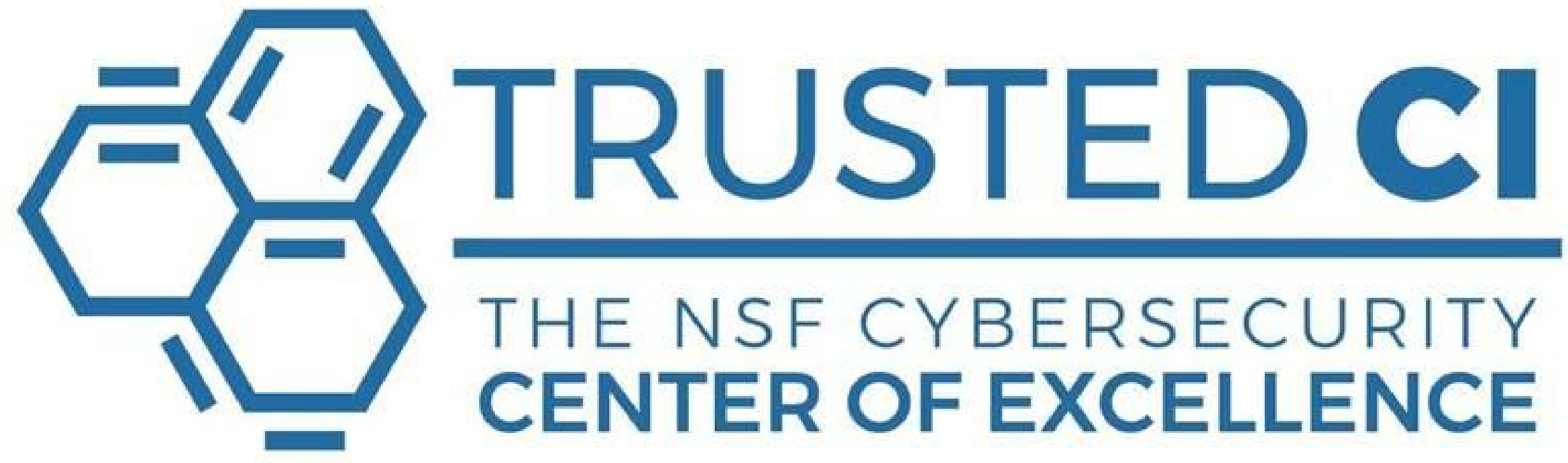

# A Study of the Reliability of Agentic AI-Generated Programs

Ayesha Shafique, Barton P. MIller, Elisa R. Heymann

**September 2026**

## About Trusted CI

The mission of Trusted CI is to provide the National Science Foundation (NSF) community with a coherent understanding of cybersecurity, its importance to computational science, and what is needed to achieve and maintain an appropriate cybersecurity program.

## Acknowledgments

This material is based upon activities supported by the National Science Foundation under Interagency Agreement #A2407-049-089-064206.0. Any opinions, findings, and conclusions or recommendations expressed are those of the author(s) and do not necessarily reflect the views of the National Science Foundation. For more information about Trusted CI, please visit https://www.trustedci.org/.

## Using & Citing this Work

# A Study of the Reliability of Agentic AI-Generated Programs

Ayesha Shafique
ashafique@cs.wisc.edu

Barton P. MIller
bart@cs.wisc.edu

Elisa R. Heymann
elisa@cs.wisc.edu

Computer Sciences Department
University of Wisconsin-Madison
Madison, WI 53706 USA

## Abstract

Agentic-AI based software development offers the promise of faster completion of the software, greater programmer efficiency, and more reliable code. The question is how can we verify these claims in an objective way? In this project, we attempted to answer this question based on three practices. First, we applied a typical best-practices agentic AI workflow for software development. Second, our target programs were ten well-known, release-quality human-written Linux utility programs so that we could compare the AI-generated code against a concrete ground truth. Third, we based our measure of reliability on a widely used testing technique, fuzz random testing. For this testing, we used both classic black box, generational testing and more modern coverage guided (gray box, mutational) testing using AFL++.

We found that the AI-generated versions of the utility programs were typically as reliable – often more reliable – than the latest human-generated versions of these programs. While the AI-generated versions did have some failures, they were less common than the code from the standard repositories. Interestingly, the AI-generated code was less likely to have failures such as memory errors (such as buffer overflows) but more likely to have hangs such as infinite loops.

In addition, we verified that generating robust and reliable software using agentic AI requires careful practice and human supervision. The quality of the code is highly dependent on the prompts and skills used, and how the human directing the process responds.

We also demonstrated that using agentic AI workflow for software development (with its prompts and skills) can become a specification of the code that leads to *cost-effective sustainability of the software*.

## 1 Introduction

With current agentic AI systems, programming has undergone a fundamental change. A significant body of code is now being generated by agentic AI systems, under the direction of a programmer. This leads to the critical question of how much should we trust this code and should we treat this code differently than the presumably human-generated code from the standard repositories?

The goal of this effort was to perform a series of experiments where we measured the reliability of AI-generated code by a well-understood and objective measure (fuzz random testing) and compare the results against comparable existing human-generated code.

We started by developing new AI-generated implementations of ten standard Linux utility programs, including `dash`, `make`, `grep`, `less`, and `tnftp`. We used Claude Code Opus 4.8 and best-practices agentic workflows based on an analysis of existing tools to specify, design, code, and test these programs [SCM+26]. As expected with AI-generated code, this process went quite quickly compared to past human-generated coding efforts. We tested these applications with classic generational black box fuzz testing using the original command line fuzz tools [MFS90][MZH22] and coverage guided (mutational gray box) fuzz testing using AFL++ [FME+20]. We then compared these results to the same tests conducted on the latest (human written) versions of these programs from the standard repositories.

There were some interesting outcomes from our study. First, as has been shown before [ROY25][SJS25], generating robust and reliable software using agentic AI requires careful practice and human supervision. The quality of the code is highly dependent on the prompts and skills used, and how the human directing the process responds. Without such care, AI systems can hallucinate and even lie. As part of this effort, we have developed a collection of prompts and skills that we used uniformly across the ten programs that we created.

Second, the AI-generated versions of the utility programs were typically as reliable – often more reliable – than the latest human-generated versions of these programs. While the AI-generated versions did have some failures, they were less common than the code from the standard repositories. Interestingly, the AI-generated code was less likely to have failures such as memory errors (such as buffer overflows) but more likely to have hangs such as infinite loops.

Third, we observed how the agentic workflow, with its prompts and skills can become a specification of the code that leads to *cost-effective sustainability of the code*. We experimented with this possibility by taking one of our AI-generated utility programs and showing how it would simply and quickly be accomplished purely by updating the AI prompts.

This study takes significant steps beyond previous research into the reliability of AI-generated code. Most of the previous research is based on simple prompts for coding, not using an agentic workflow. For example, Pearce et al [PAT+22] used GitHub Copilot inline suggestions for autocompleting programs, not agents. They concluded that ~40% of the generated code contained vulnerabilities. Similarly, in Perry et al [PSK+23], a group of programmers were required to write short programs, and were allowed to query an AI assistant. The resulting code was then compared to that generated by programmers who did not have access to AI tools. The researchers found that participants with AI assistance wrote significantly less secure code, yet were more likely to believe their code was secure. Additionally, in Shukla et al [SJS25], the code, usually very short programs, was generated by iteratively rewriting existing seed programs. They found that iterative refinement tended to degrade security; even security-focused prompting did not prevent vulnerability accumulation across iterations.

There were a few studies that did assess the security of agentic-AI generated code. One recent project [CHL+26] used agents to generate code patches to implement new features in an existing code base. That project focused solely on memory-safety issues. The security of the patch was determined by running known exploits for the code, comparing the results for the versions before and after the patching. They found that the best agent produced correct and secure solutions only 23.8% of the time.

Our paper makes contributions based on four pillars:

1. We evaluated the reliability of AI-generated code produced by a typical best-practices agentic AI workflow. This is in contrast to simple prompt coding (often called "vibe coding").
2. We compared the quality of the AI-generated code against a concrete ground truth, where the ground truth is release-quality human-written code.
3. We used a well accepted measure of reliability, fuzz random testing [MFS90], using both classic black box and gray box coverage guided testing.

4. We showed how the AI prompts for a specification for the program can easily be updated, thus allowing software to be sustained by new developers without requiring almost any knowledge of the code.

In Section 2, we first discuss our agentic AI-based software development methodology, including the processes for design, coding, and testing. In Section 3, we present the list of current human-written Linux utility programs that we used as our ground truth for calibrating the reliability of the AI-generated versions. Section 4 reports on our experiences generated the AI versions of the utilities. Section 5 describes how we used both classic and coverage guided fuzz testing to evaluate both the human- and AI-generated versions of the programs. We present the testing results in Section 6 and then discuss AI-based software sustainability in Section 7. Section 8 is a discussion of the results and concludes our work.

## 2 Agentic AI Code Development Methodology

Our goal in developing the AI-generated versions of the Linux utilities was to follow current best practices in agentic AI-based software development, meaning that we followed a methodology similar to one that would be used by an experienced software engineer when leading a team of "virtual developers". This decomposition dates back to early work on agent-based software engineering, which similarly treats development as the coordinated management of multiple specialized agents, with responsibilities divided across specification, implementation, and verification [Woo97]. First, just as a well-managed software project begins from an explicit statement of its required behavior [Roy70], our methodology treats the utility's documentation as the authoritative specification. Second, just as software projects use coding standards and structured review practices to improve consistency and reduce defects [Fag76], our methodology uses custom, reusable instructions and rules to impose a common development discipline across agents. The rules standardize decisions about language, coding style, testing, and review, rather than leaving them to ad hoc prompting in each session. Third, just as mature software teams separate design, implementation, and review [Fag76], our workflow separates these responsibilities across agents. Figure 1 illustrates our workflow pipeline.

It is important to note that an agentic AI-enabled software workflow is not what is colloquially known as "vibe coding". This is not just writing a prompt like "write me a quicksort program in Python". Agentic AI-enabled software development is like managing a software development team, where your team members are AI agents of various types. Like classic software development, it requires rigor, well-defined practices, and constant innovation. In this section, we describe the practices that we used. Our goal was not to innovate on these practices, but to capture and model practices that have been found to be effective, and compare them to human written code.

We now describe the methodology in more detail. Section 2.1 explains how the utility documentation is used as the only behavioral specification. Section 2.2 describes the custom skills used to control the workflow. Section 2.3 presents the five-stage agentic pipeline used to derive the test oracle, design the program, implement it, and audit the resulting system by performing an evaluation of the code.

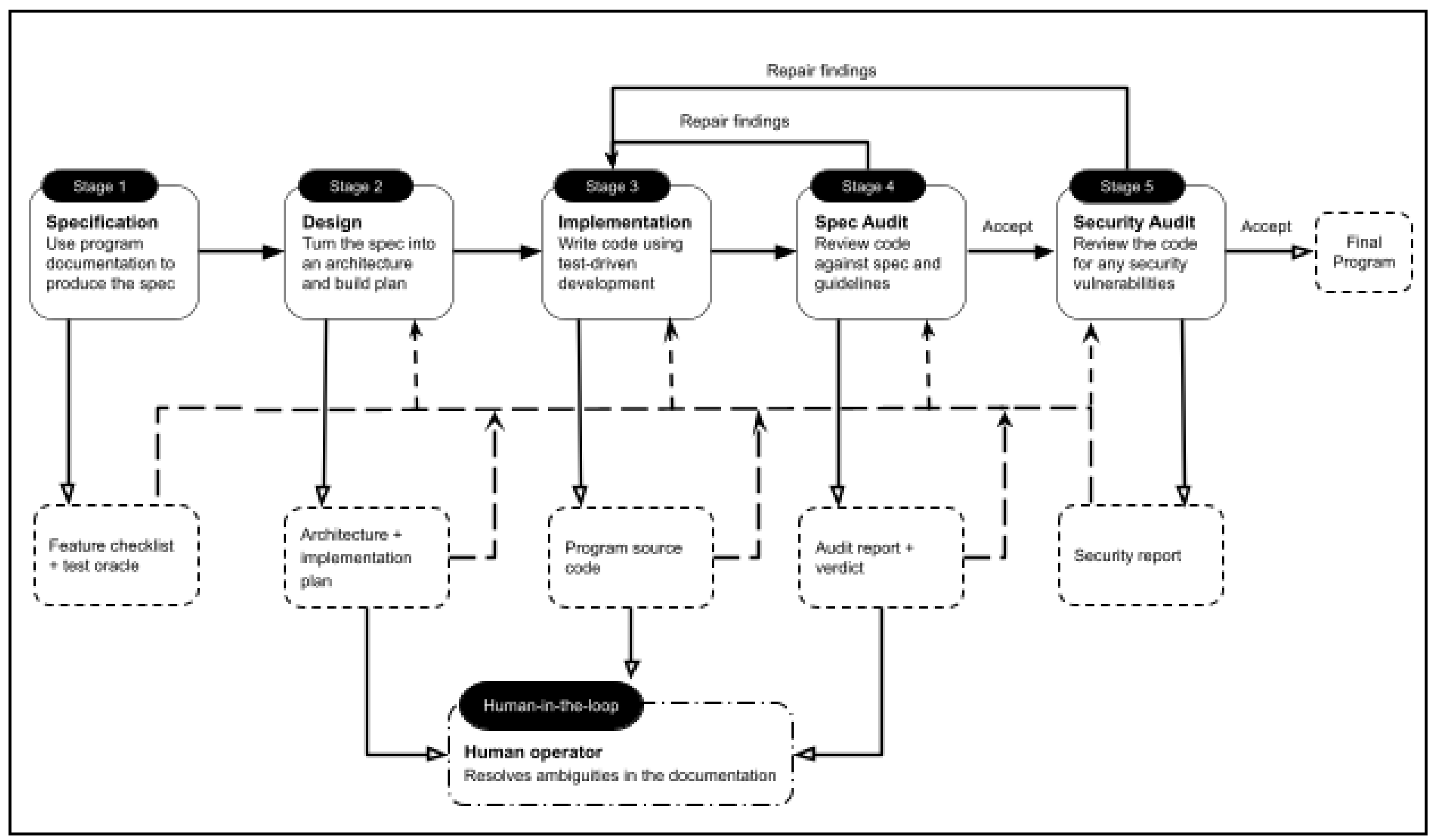

**Figure 1. Agentic AI Software Development Workflow Pipeline**

## 2.1 The Utility Program User Documentation is the Only Specification

The goal of our study is to model AI-based software development. As such, we want the development to be independent of the currently available versions of the utility programs. Each program is built against the authoritative specification, which is based on the published manual page for the utility, along with other more detailed user manuals such as those found for utilities such as `grep`. We do not allow reference to or execution of the installed binary version of the program or the repositories that contain the sources for the installed version. We do note that previous training activities may have allowed the AI system to incorporate some knowledge of the selected utility programs. Details of such knowledge are difficult to track. We do note from our inspection of the AI-generated code and the repository versions of the code, that they do not show noticeable similarities.

We consider the program to be correct if it implements the specification; we do not require the output to be character-by-character identical to the original repository version.

Our development methodology may differ from other AI-driven programming projects in that we get to start with existing documentation on the programs' functionality. However, even with the well-established documentation for the utilities that we built, there are ambiguities and incomplete specifications. Such issues are detected during testing and require iteration through the development stages. This need for iteration of the design is what is expected in any software development project.

## 2.2 Custom Skills vs. Marketplace Plugins

We wrote custom skills to define the rules of the workflow. A skill is a markdown document that tells an agent what to read, what to produce, and which rules to follow [XY26]. When an agent invokes the skill, its complete instructions are added to the context and guide the agent's subsequent work. We tailored each of our skills to the class of programs that we were developing and to our specific workflow. This followed Anthropic's guidance that skills should encode instructions and expertise for specific tasks and repeatable workflows pertaining to a project [Ant25b]. We built the skills using Anthropic's skill-creator tool [Ant25a].

The alternative to developing custom skills was to use general-purpose plugins from Anthropic's marketplace such as Superpowers [Vin26], which presents a complete software-development methodology composed of reusable skills. Such plugins are designed to support a broad range of projects, but our workflow required more specialized instructions for reconstructing Linux utilities in C from their documentation. Including instructions that are not relevant to the current task can consume limited context and distract the model from the information that matters most. Anthropic's context-engineering guidance therefore recommends keeping an agent's context focused on information that is directly relevant to the task [Ant25c]. This recommendation is supported by empirical studies showing that irrelevant information can distract language models, reduce task accuracy, and influence generated code in incorrect ways [JS22][SCD+23]. We instead wrote narrowly scoped skills that encoded our exact requirements, such as implementing the utilities in C and treating the manual page as the behavioral contract.

## 2.3 Our Agents and Workflow

The workflow runs in five stages, orchestrated by a main agent. Each stage corresponds to a distinct phase of the development process and has one clearly defined responsibility. This decomposition separates concerns across role-specialized stages. Recent multi-agent software engineering frameworks divide work in the same way, using role-specialized phases and intermediate artifacts to structure collaboration, reduce cascading errors, and support independent verification [HZC+24][QLL+24][WBZ+24]. This staged decomposition structures the workflow at the level of the overall development process. Within each stage, the main agent applies the same principle at a finer level by dividing its responsibility into smaller, bounded tasks and assigning them to fresh subagents. In this subagent-driven development pattern, each subagent receives a narrow objective while the main agent consolidates the results of each subagent. We decompose the work this way for three reasons. First, a language model performs better on a task when it is split into smaller, clearly specified problems that are each solved on their own [KTF+23]. Second, every subagent starts from a clean context that contains only its own task. It never sees the leftover state of the tasks that came before it, so an early mistake or an irrelevant detail cannot accumulate in its context and affect its output as the work progresses [LLH+24]. Third, sub-tasks that do not depend on one another can run in parallel via different subagents, reducing development time [Ant25c].

The main agent ran on the strongest model available to us at the time of writing (Opus 4.8), with a one-million-token context window and an extra-high thinking budget. We selected the extra-high setting because the tasks required substantial reasoning over large code bases, specifications, and test results. In our experiments, reducing the thinking budget produced a noticeable degradation in performance, particularly for the more complex implementation and debugging tasks. For the simpler and more mechanical work, the main agent launched subagents on cheaper models. For example, the main agent launches a subagent on the smaller model (Haiku) to fetch the manual page.

### 2.3.1 Stage 1: Derive the Feature Checklist and the Test Oracle

Our workflow starts with Stage 1 where we define what the program is expected to do. This expected behavior is derived from the program's specification, as described in Section 2.1. The purpose of this stage is to convert the specification into two explicit artifacts, a feature checklist and a test oracle.

The feature checklist is a list of the documented behavior that the agent must implement, with one entry for each documented behavior. The test oracle is the suite of test cases used to guide test-driven development during implementation and check whether the agent-written code conforms to the specification.

The main agent for this stage produces these artifacts through two skills. The first skill reads the specification and extracts the feature checklist. We use a feature checklist rather than leaving the specification as prose because an agent works more effectively from an explicit list of requirements. When requirements are listed individually, the agent is better able both to produce output that satisfies them and to check whether the resulting output meets them [WWQ+25][VSM+25]. The second skill converts each checklist item into test cases. This process follows specification-based testing, in which tests are derived from the stated behavior of a program rather than from its source implementation [OB88]. Prior work has also shown that manual pages and other program documentation can provide input constraints, expected behavior, and test oracles for generated tests [WZW+15]. When publicly available tests exist for the utility, we also incorporate them alongside the test suite generated by the agent. We include these tests because existing test suites often capture edge cases and long-standing behavioral assumptions that can be difficult to extract completely from the documentation alone.

### 2.3.2 Stage 2: Design the Program

Stage 2 establishes the design of the program. Its main agent works from the feature checklist produced in Stage 1. We fix the design and implementation plan before writing any code so that the agent resolves the program structure and any gaps in the specification in advance. The resulting plan gives the agent that writes the code a stable reference to follow, rather than requiring it to make architectural decisions while coding. Prior work shows that a language model will reason more reliably and generate better code when it plans a task before implementing it [JDW+24].

The main agent for this stage produces the design through two skills, each of which produces a written artifact. The first skill produces the program architecture. This document describes the high-level structure of the program. It identifies the main components, the major data structures, error behavior, resource management, and any special mechanisms the program needs, such as terminal handling or concurrency. The second skill produces an implementation plan. This names the modules the program will be built from and records every place where the specification is silent or ambiguous. These cases are escalated to the human operator, together with the agent's suggested resolutions. The operator specifies the resolution to adopt, and that decision becomes part of the implementation plan. After producing both documents, the main agent launches a critic subagent to review them against the specification and feature checklist. The critic identifies uncovered requirements, contradictions with the specification, and weaknesses in the proposed design. It returns actionable feedback to the main agent, which revises the relevant document. This review-and-revision cycle continues until the critic raises no further issues. Prior work shows that language-model outputs can improve through repeated feedback and revision [MTG+23], and that feedback from a separate critic can provide specific, actionable guidance that helps an agent make better decisions [YFL+25].

When a question requires background research, the main agent or critic may launch research subagents to check documentation, summarize relevant architecture patterns, or gather technical background.

### 2.3.3 Stage 3: Write the Code

Stage 3 implements the program in C. The main agent for this stage works from the architecture and implementation plan produced in Stage 2, together with the feature checklist and test oracle produced in Stage 1. This means that code generation begins only after the expected behavior has been extracted, the tests have been defined, and the design has been written down.

The main agent produces the program implementation by using three skills. The first skill defines the coding process. Instead of asking the agent to write the whole program at once, this skill makes the agent implement the program one checklist item at a time from the feature checklist [KTF+23]. Each checklist item becomes a small, bounded coding task. This structure makes the implementation process more efficient and manageable in two ways. First, the agent only has to reason about one documented behavior at a time, rather than the full program at once. Second, when two checklist items do not depend on each other, the main agent can assign them to separate fresh subagents that work in parallel. Each subagent receives only the instructions, code, and checklist item needed for its assigned task. This keeps the subagent's context short and focused, reducing the chance that relevant information is lost among unrelated details [LLH+24]. For each checklist item, the agent writes the code, runs the test oracle, finds the behavior that still fails, applies the fixes, and runs the oracle again. This process provides external execution feedback rather than requiring the agent to identify errors solely by reviewing its own output, a form of intrinsic self-correction that prior work has found to be unreliable [HCM+24]. Prior work also shows that execution feedback improves language-model code generation and debugging [CLSZ24]. For each checklist item, the agent repeats the same test-driven cycle. The cycle ends only when the full test oracle passes, and every checklist item has either been implemented or explicitly deferred. A checklist item is deferred only when it does not apply to the target Unix environment, such as behavior documented for another platform.

The other two skills define coding standards that are to be followed by the agent throughout implementation. One skill covers defensive programming, validation at trust boundaries, secure memory handling, and the SEI CERT C rules [SEI16]. This explicit security guidance is important because prior work shows that prompting techniques tailored for secure code generation can reduce security weaknesses in LLM-generated code [TDM+25]. The third skill covers clarity, modularity, maintainability, and code style. These skills make reliability, security, and maintainability part of the coding process of the agent.

### 2.3.4 Stage 4: Audit the Program

Stage 4 performs an independent conformance audit of the completed program. Here, a conformance audit is a structured review that determines whether the implementation satisfies the documented requirements and behaves as intended. The main agent for this stage runs in a clean context, separate from the agent that generated the code, thereby decoupling evaluation from implementation. This separation is consistent with prior work showing that interactions among multiple model instances can improve the correctness and factual accuracy of their tasks [DLT+24].

The main agent conducts the audit through one skill. The agent reviews the source code, builds and executes the program on selected inputs, and evaluates it against the specification, the Stage 1 feature checklist and test oracle, and the Stage 2 implementation and architecture plan. It identifies missing functionality, behavior that diverges from the specification, incorrect handling of documented cases, unsafe assumptions, and observable memory errors. These explicit criteria are motivated by prior work showing that language models can evaluate outputs effectively when given clearly stated requirements [ZCS+23].

The audit returns one of three verdicts. ACCEPT indicates that no missing features or specification divergences were found. KICK BACK indicates that a required behavior is missing or incorrectly implemented; the audit findings are returned to the Stage 3 agent, which repairs the program before Stage 4 audits it again. This cycle continues until the audit returns ACCEPT, following prior work showing that iterative critique and revision can improve language-model outputs [MTG+23]. ESCALATE indicates a genuine ambiguity in the specification that requires a decision from the human operator.

The program is considered complete only when it builds cleanly, passes the full Stage 1 test oracle, accounts for every item in the Stage 1 feature checklist through implementation or explicit deferral, and receives an ACCEPT verdict.

### 2.3.5 Stage 5: Security and Correctness Audit

Stage 5 examines the completed program for security defects that may remain even when it behaves according to its specification. The main agent for this stage runs as a fresh, independent agent in a clean context and follows a dedicated skill for systematically auditing C code. This skill directs the agent to inspect the source and, where appropriate, use compiler diagnostics, static analyzers, sanitizers, and debugging tools such as GDB and Valgrind to support its security analysis. These are the same kinds of tools available to a software engineer performing a manual security review, and they give the agent additional evidence when reasoning about possible defects. The audit targets memory-safety violations, resource-management errors, unchecked inputs or return values, and other exploitable vulnerabilities. The skill also instructs the agent to produce actionable findings, including the affected code, the likely cause, and a concrete recommendation for repair. We explicitly prohibit the use of fuzz testing at this stage because we reserve it for the independent evaluation of the completed programs. If we allowed the agent to fuzz its own implementation, it would expose the agent to the same technique later used to measure its security, compromising the independence of the evaluation and potentially biasing the results in the agent's favor.

Stage 5 begins only after the Stage 4 audit has been completed and accepted, meaning that the program has already been verified against the specification, feature checklist, and test oracle. Its purpose is therefore not to reassess conformance to the documentation, but to identify implementation-level security and code-quality defects that may remain despite correct documented behavior. Recent work by Anthropic motivates this step by showing that security-focused coding agents can identify and validate vulnerabilities in well-tested code bases and produce actionable recommendations for repair [C+26].

The main agent may launch and coordinate multiple subagents to parallelize the audit across different parts of the program or classes of defects. At the end of each audit, it produces a security audit report listing the findings, the supporting evidence, and, where possible, a recommended fix. Each finding is assigned a severity level of HIGH, MEDIUM, or LOW. LOW-severity findings primarily concern code quality, such as dead code, unnecessary complexity, or minor maintainability problems, rather than significant security risks. The report is returned to the Stage 3 agent, which fixes the reported bugs. The revised program then passes through Stage 4 before Stage 5 performs another security audit. This cycle continues until the report contains no HIGH- or MEDIUM-severity findings or the process reaches a maximum of 20 audits.

## 3 The Ground Truth – Human Written Utilities

We selected 10 existing Linux utility programs as our ground truth. These programs were selected because of our long history of using UNIX utilities in our fuzz testing research and the broad community familiarity with them. We used the latest published release version of each utility that was found in its public repository. We compare the system utilities version vs the latest public version we used in the table for each of our ten applications

They vary from around 400 lines of code up to 18,000 lines. All the programs are written in C and are executed from the command line. Table 1 lists the utility programs tested, the versions tested, line counts for both versions, and the repository location.

### 3.1 Code Size Measurement

The programs that we chose are in common use, have current implementations, and vary in complexity. Calculating the number of source lines in a program can be surprisingly complex. To compare the size of the repository and AI-generated implementations, we measured only the source code that contributed to each final executable. Counting an entire source tree can overestimate program size because packages often contain shared or unused code. We therefore counted only non-blank, non-comment lines from C functions that were actually called and global variables actually used.

To identify the actually-used code for line counting purposes, we rebuilt each program so that individual functions and data objects could be removed independently by the linker. Optimization and inlining were disabled, while linker garbage collection removed unused definitions. We then examined the executable's symbol table to determine which definitions remained.

We mapped those definitions back to their source code using Clang. After ensuring consistent formatting, we excluded blank lines and comment-only lines and counted only code active on the Linux system used for our experiments. Lines containing both code and comments were counted as code, and the final counts were obtained using `cloc`.

We included C source and header files, including generated C code when it contributed to the executable. We excluded build scripts, documentation, assembly files, dynamically linked libraries, the standard C library, gnulib regex code, startup code, and source lines outside retained functions or global variables. We also excluded source lines outside retained functions or global variables, such as include directives, function prototypes, type definitions, and macro definitions. Macro uses within retained code were still counted as code.

We note that any line counting mechanism is going to be imperfect because of the complexities of real code repositories. Our current methodology represents our best current approach.

## 4 AI Generated Utilities

This section reports our experience applying the five-stage workflow to the selected utilities. Stages 1 and 2 generally proceeded as intended and did not reveal recurring problems that required changes to the workflow. The most notable issues arose during test-driven implementation in Stage 3 and during the specification conformance and security audits in Stages 4 and 5.

| Program | Repo Version | LOC (using cloc) | | LOC ratio Repo : AI |
|---|---|---|---|---|
| | | Repo | AI | |
| `cat` | 9.4 | 1,583 | 439 | 3.61 |
| | `https://github.com/coreutils/coreutils` | | | |
| `dash` | 0.5.13.4 | 10,416 | 12,662 | 0.82 |
| | `https://git.kernel.org/pub/scm/utils/dash/dash.git` | | | |
| `dc` | 1.5.2 | 2,809 | 2,572 | 1.09 |
| | `https://ftp.genu.org/gnu/bc/` | | | |
| `grep` | 3.12 | 8,089 | 4,428 | 1.83 |
| | `https://ftp.gnu.org/gnu/grep/` | | | |
| `less` | 702 | 17,517 | 19,582 | 0.89 |
| | `https://github.com/gwsw/less` | | | |
| `make` | 4.4 | 15,158 | 14,378 | 1.05 |
| | `https://ftp.gnu.org/gnu/make/` | | | |
| `ptx` | 9.11 | 3,123 | 1,582 | 1.97 |
| | `https://github.com/coreutils/coreutils` | | | |
| `strings` | 2.46.1 | 991 | 889 | 1.11 |
| | https://ftp.gnu.org/gnu/binutils/ | | | |
| `tac` | 9.11 | 2,140 | 412 | 5.19 |
| | `https://github.com/coreutils/coreutils` | | | |
| `tnftp` | 20260211 | 9,752 | 9,026 | 1.08 |
| | `https://ftp.netbsd.org/pub/NetBSD/misc/tnftp/` | | | |

**Table 1: The Utility Programs Tested**

## 4.1 Stage 1 Experiences

Stage 1 proceeded without significant difficulty for all utilities. The main challenge was the amount of documentation that had to be processed for some programs. Several utilities had specifications extending well beyond a short manual page. For example, the documentation used for `grep` comprises approximately 17 pages, while utilities such as `less` have an unusually extensive manual page and GNU `make` has a separate, comprehensive GNU manual covering its full language and behavior. Despite the size of these specifications, the agents were generally able to extract the documented behavior into a feature checklist and corresponding test oracle.

## 4.2 Stage 2 Experiences

Stage 2 also proceeded smoothly for most utilities. The agents generally produced reasonable architectures, data structures, and error-handling strategies, and these designs rarely required major revision. Most of the effort instead went into resolving ambiguities in the documentation, especially for larger utilities where such decisions formed a substantial part of the implementation plan.

However, many important ambiguities only became apparent during implementation, when features interacted or concrete design choices had to be made. Stage 2 therefore helped resolve ambiguities that were visible in advance, but could not eliminate those that emerged only through implementation. Moreover, identifying a requirement in the design did not guarantee that it would be implemented. For example, the plan for `less` correctly identified incremental scrolling as necessary for several documented options, but the scrolling mechanism was never implemented, leaving those options accepted but ineffective.

Overall, Stage 2 was useful for recording design decisions and resolving early ambiguities, but its plans were neither necessarily complete nor always followed faithfully during implementation.

## 4.3 Stage 3 Experiences

Stage 3 was the most time-consuming stage of the workflow. For the smaller utilities, the agents were generally able to implement features, run the tests, and make corrections quickly. This changed as the programs became larger. For utilities such as `less` and `make`, there were many iterations of this stage, requiring hundreds of commits.

Several factors contributed to this behavior. One problem was determining the correct behavior of the program. We used three sources: the documentation, the source repository test suite, and the behavior of the original utility. These sources did not always agree. We found cases where the manual contradicted itself, where the repository tests required behavior that was not documented, and where the original program behaved differently from the manual. The agents were not allowed to inspect or execute the reference implementation. When an ambiguity could not be resolved from the documentation or tests, the human operator could run the same input on the original utility and use that result to decide which behavior to follow.

A second problem was that the agents' own reports of progress were often unreliable. We observed cases where an agent declared a feature complete even though it had only been stubbed or partially implemented. A common example was a command line option that was parsed and stored but never used by the rest of the program. In `less`, eleven such command line options were parsed but never used. We also found that agents would often claim that a feature had been waived, and hence skip implementing it, even though no such waiver appeared in the plan or had been approved by the human operator. This suggests that claims produced by an agent should be treated as hypotheses to verify, rather than as the truth.

The agent-generated test suites could also give a misleading impression of completeness. The agents did not always generate tests that exercised the full behavior of a feature. In some cases, a test only checked that a command line option was accepted, without verifying that the option actually affected the program's output or behavior. The eleven unused command line options in `less` survived four audit rounds and a test suite in which 1217 of 1221 tests passed. This made manual testing and repository test suites important independent checks. One improvement to the workflow would be to add a separate review of the test suite itself. This stage could classify tests by how strongly they exercise the corresponding feature, distinguishing tests that only check parsing or successful execution from tests that verify observable behavior and edge cases. Such a step would make weak tests easier to identify before they are used as evidence that a feature is complete.

`Make` provided the clearest example of this issue. Although all of the agent-generated tests passed, the resulting program could initially build little beyond a simple C program. We therefore tested it by building real software. This software included Linux kernel builds using tiny, default, and full configurations, together with 17 other software packages, including PostgreSQL, 7-Zip, Redis, BusyBox, SQLite, FFmpeg, OpenSSL, binutils, and curl. These builds exposed multiple missing and incomplete features that the agent-generated tests had not detected and required approximately another week of implementation and debugging. For a program such as `make`, successfully building real software provided a substantially stronger measure of completeness than passing tests derived from the same specification used to generate the implementation.

As the programs became larger, implementation also required repeated cycles of regression testing and repair. Changes made to implement or correct one feature often caused previously working behavior to fail. The agents generally detected these regressions when the existing tests exercised the affected behavior and then repaired them in subsequent iterations. For the larger utilities, this resulted in multiple rounds of implementation, testing, and correction before the program stabilized.

The agents' algorithmic and design choices did not consistently match the quality of the repository implementations. In `ptx`, the agents used an unnecessarily expensive algorithm with quadratic behavior, while the repository version used a linear-time approach. In contrast, the `tnftp` implementation was more modular than the repository version and added safeguards such as better state isolation and a recursion-depth limit. These cases show that the agents could make both weaker and stronger design choices than the repository implementations.

Overall, Stage 3 showed that the limiting factor for large agent-generated programs was not code generation itself, but verification. As the programs became more feature-rich, confidence increasingly depended on evidence that was independent of the implementing agent: upstream tests, manual testing, realistic workloads, and repeated review.

## 4.4 Stage 4 Experiences

Stage 4 also became substantially more time-consuming as the utilities increased in size. Smaller programs such as `ptx` and `tac` generally required only a few audit rounds, while larger utilities required many rounds before they were accepted. For example, make ultimately required more than 20 Stage 4 rounds as successive audits continued to identify substantive findings.

A single audit round was often not sufficient. Later rounds found defects that earlier rounds had missed, as well as defects introduced while fixing previous findings. For instance, in `grep`, a missing implementation of a command-line option documented in the manual survived four earlier audit rounds, the agent-generated tests, and the repository test suite before it was detected. This suggests that an ACCEPT verdict from audits should not be treated as strong evidence that a program is complete.

The audit findings themselves also required verification. Several rounds reported issues that did not exist. In grep, an audit incorrectly concluded that a negative argument to the command line option limiting the number of matching lines should be rejected, even though the manual defines this value as meaning unlimited. A fix was applied, but the next audit re-read the manual, identified the change as a regression, and reverted it. This showed that, although individual audit rounds could produce incorrect findings, subsequent independent rounds were often able to correctly identify and correct those mistakes.

Repeated audits were also not sufficient by themselves. make passed its agent-generated tests and more than 20 audit rounds, yet the resulting program could initially build little beyond a simple C program. The missing functionality became apparent only when we used it to build the Linux kernel and other real software, as described in Stage 3. Thus, repeated auditing was effective at finding many local defects, but it could not establish that a complex utility worked correctly in realistic use or identify every missing or partially implemented feature described by the documentation. For the larger utilities, audits therefore had to be combined with manual testing, repository test suites, and realistic end-to-end workloads.

## 4.5 Stage 5 Experiences

Like Stages 3 and 4, Stage 5 became substantially more time-consuming as the utilities increased in size. The smallest utilities required only one or two rounds before the agents stopped reporting HIGH- or MEDIUM-severity findings, while larger programs required substantially more. make, for example, required fourteen Stage 5 rounds before reaching this point.

As described in Section 2.3.5, the agents were allowed to use compiler diagnostics, static analyzers, sanitizers, gdb, Valgrind, and other standard debugging tools during their security analysis. We found that Claude Code agents could use these tools effectively without additional guidance from the human operator. Runtime tools such as sanitizers and Valgrind, however, can expose a defect only when an input exercises the affected code path. Finding such inputs is difficult for command-line utilities because they support many flags, modes, input formats, and combinations of program state. The agents addressed this problem by using the source code to guide their testing. They examined the code for suspicious operations and failure paths, inferred the conditions needed to reach them, and constructed targeted inputs to exercise those paths under appropriate analysis tools. Once an input triggered a defect, the tools provided detailed evidence about the failure, such as the location of an invalid memory access, arithmetic overflow, or use of uninitialized memory.

This approach was particularly effective. For instance, in `less` source inspection and targeted testing led the agent to inputs for which the undefined-behavior sanitizer reported arithmetic overflows, the memory sanitizer detected out-of-bounds reads, and Valgrind identified more than 200 uses of uninitialized memory.

Another example was `dash` which used `siglongjmp` to recover from syntax and evaluation errors. During code review, the auditing agents found state changes and resource acquisitions whose restoration code such a jump could bypass. One example involved the lexical analyzer. When a syntax error occurred while processing an input containing a NULL byte, `siglongjmp` bypassed the function that reset the lexical analysis state. The shell then retained stale state and repeatedly processed the same invalid input, resulting in an infinite loop. The agents identified the problem by tracing the recovery path through the source, constructed targeted NULL-containing inputs to reproduce it, and successfully repaired the reset logic.

The agents also looked beyond bugs that could directly crash or compromise the utility itself. They considered whether the utility could be used as a path to attack the user. For example, while auditing `tnftp`, an agent noticed that replies from a remote FTP server could be printed directly to the user's terminal. A malicious server could include ANSI escape sequences in those replies, causing the terminal to interpret them as control commands instead of ordinary text. This showed that the agents considered not only how the program itself could be attacked, but also how an attacker could use the program to affect others.

We also observed that agents did not always find all instances of the same defect pattern in a single round. In `less`, one round fixed a file-descriptor leak, the next found three additional paths with the same problem, and another round two rounds later found the final case. This suggests that once an audit identifies a defect pattern, it should explicitly examine similar code paths rather than assume that the first instance is isolated.

After each Stage 5 audit, the Stage 3 agent repaired the reported defects. These repairs could introduce regressions, while later Stage 5 rounds could also uncover security defects that earlier rounds had missed. We therefore sent every repaired program through Stage 4 again before beginning another Stage 5 audit. This allowed the workflow to check that the repair had not broken previously correct behavior before continuing the security analysis.

Overall, Stage 5 showed that the agents could perform much of the security-audit and repair process without direct assistance from the human operator. They independently examined the source, identified suspicious code paths, constructed targeted inputs, selected appropriate analysis tools, interpreted their diagnostics, and repaired many of the defects they found. The analysis tools provided important evidence, but the more significant observation was that the agents could combine these tools with source-level reasoning to direct the investigation themselves.

## 5 Testing Methodology

To provide a concrete and objective measure of the reliability of the utility programs, we used fuzz testing, both traditional black box testing and more modern gray box coverage guided testing. While fuzz testing is not a complete testing methodology, it is well understood and accepted, is good at finding inputs that will challenge the assumptions in the code, and has a long history for finding bugs in programs. It also provides a uniform testing strategy across a variety of different programs.

### 5.1 Traditional Black Box Fuzz Testing

Black box, generational, and unstructured random testing was the original type of fuzz testing [MFS90]. It is easy to use, in part due to its unstructured input and simple test oracle that only considers the program to have failed if it crashed or hung. While there has been an enormous amount of progress in fuzz testing since this was introduced, recent research has shown it to still be an effective technique [MZH22]. For this testing, we used the original fuzz tools, as updated for modern versions of C.

We generated eight sets of random inputs covering four size ranges, from inputs smaller than 1 KB to inputs as large as 100 MB. In total, these inputs contained 4,280 test cases and approximately 12 GB of data. The inputs varied in character content, including arbitrary bytes, printable characters, and NULL bytes, and were generated both as unstructured byte streams and as line-oriented input. This random data was supplied through standard input, input files specified as parameters, and, for interactive utilities, a pseudo-terminal. For interactive programs, an appropriate quit sequence was appended where possible. We also varied the command-line options, with each option from a configured pool independently selected with probability 0.5.

As is common with fuzz testing, the main difficulty is distinguishing genuine hangs from programs that were simply waiting for more input or taking a long time to process very large test cases. Testing interactive programs such as `dash` and `less` also created the possibility of side effects outside the target program. Random input could modify shell configuration or history files and, more importantly, could accidentally invoke commands or launch other processes. For example, a shell could execute a command such as yes, which would run indefinitely, or launch an editor or another interactive program that would then wait for additional input. Other random command sequences could start background processes that continued running after the test completed, leaving behind resource-consuming or zombie processes. Random shell input could even invoke system-management commands such as a reboot. Programs such as `less` could similarly launch external commands through its interactive interface. We therefore ran such utilities inside Docker containers to isolate these effects and manually examined reported hangs before classifying them as failures.

## 5.2 Coverage Guided Fuzz Testing with AFL++

Modern fuzz testing tools have made significant advances over the original black box tools. Two important advances are: (1) mutational, in that they initial “seed” inputs and then randomly modify these inputs to generate news ones; (2) gray box, in that they track which basic blocks execute to identify which test inputs should be considered for further mutation. These tools can be somewhat more complicated to use as they require a special compiler that will insert instrumentation to track which basic blocks execute. For this testing, we used the popular and effective AFL++ tool [FME+20].

We compiled each utility with `afl-clang-fast` and tested it using AFL++ version 5.01a inside Docker containers. For each utility, we created a small seed corpus containing representative valid inputs together with edge cases such as empty, malformed, binary, and unusually long inputs. Where appropriate, we also supplied dictionaries containing syntax commonly recognized by the utility, such as shell operators for `dash`. AFL++ uses these dictionary entries during mutation to increase the likelihood of generating inputs containing syntactically meaningful tokens [AFL26].

We tested each utility in two separate campaigns because AFL++ normally mutates a single source of input at a time. In the first campaign, AFL++ mutated the data read from standard input or an input file while the utility was repeatedly invoked with no command-line options. In the second, we used AFL++'s `argv-fuzz` support to mutate the command-line arguments while the utility processed fixed input data, allowing AFL++ to vary both which arguments were supplied and their values. This approach differed from our classic fuzz testing, where random input data and command-line options were varied together, with each option selected independently with a fixed probability. The repository and AI-generated versions of each utility received the same seeds and testing configurations.

A major difficulty was determining which of the crashes and hangs reported by AFL++ represented reproducible failures. Some reports depended on timing or environmental state and could not be reproduced consistently, so we replayed the reported inputs and reduced them before counting them as distinct bugs. This process was particularly important for programs such as `dash` and `make`, which could also produce side effects by launching processes in the background or modifying files, similar to the issues encountered during classic fuzz testing. We therefore ran the campaigns inside Docker containers to isolate filesystem and process side effects and replayed reported hangs individually with longer timeouts. The campaigns themselves were also time-consuming, with several running for days while AFL++ continued discovering new execution paths. The AFL++ documentation notes that fuzzing campaigns commonly run for days or weeks and may continue cycling while new paths are still being discovered [AFL26].

| Program | | Total | | Classic | | AFL++ | |
|---|---|---|---|---|---|---|---|
| | | Crash | Hang | Crash | Hang | Crash | Hang |
| `cat` | Repo | 0 | 0 | 0 | 0 | 0 | 0 |
| | AI | 0 | 0 | 0 | 0 | 0 | 0 |
| `dash` | Repo | 2 | 0 | 1 | 0 | 1 | 0 |
| | AI | 0 | 0 | 0 | 0 | 0 | 0 |
| `dc` | Repo | 0 | 1 | 0 | 1 | 0 | 1 |
| | AI | 0 | 1 | 0 | 1 | 0 | 1 |
| `grep` | Repo | 0 | 1 | 0 | 0 | 0 | 1 |
| | AI | 0 | 1 | 0 | 0 | 0 | 1 |
| `less` | Repo | 3 | 0 | 3 | 0 | – | – |
| | AI | 0 | 0 | 0 | 0 | – | – |
| `make` | Repo | 1 | 0 | 0 | 0 | 1 | 0 |
| | AI | 0 | 0 | 0 | 0 | 0 | 0 |
| `ptx` | Repo | 1 | 1 | 0 | 1 | 1 | 1 |
| | AI | 1 | 1 | 0 | 1 | 1 | 1 |
| `strings` | Repo | 0 | 0 | 0 | 0 | 0 | 0 |
| | AI | 0 | 0 | 0 | 0 | 0 | 0 |
| `tac` | Repo | 1 | 1 | 0 | 0 | 1 | 1 |
| | AI | 0 | 0 | 0 | 0 | 0 | 0 |
| `tnftp` | Repo | 7 | 0 | 2 | 0 | 5 | 0 |
| | AI | 1 | 0 | 1 | 0 | 0 | 0 |

**Table 2: Fuzz Testing Results for Repository and AI-Generated Versions of Utilities**
*Testing was done with black-box fuzz testing using the original fuzz tools and modern coverage-guided testing using AFL++. Classic black-box fuzz testing uniquely identified 7 crashes and 4 hangs, AFL++ uniquely identified 10 crashes and 7 hangs, and both methods identified 4 common hangs, for a total of 17 unique crashes and 7 unique hangs.*

## 5.3 Debugging Failures

We debugged each failure reported by the classic and coverage guided fuzz testing to identify the root cause of the failures. The root cause was a specific bug at a specific line of code. Note that there were often multiple fuzz inputs that triggered the same bug, so these were combined into one report. The results that we present in the next section are the number of unique failures of each utility program.

## 6 Test Results

Overall, the fuzz testing exposed 24 unique failures across the ten utility programs, repository and AI-generated versions. Of these 24 failures, 19 were in the repository versions of the programs and 5 were in the AI-generated versions. Classic black-box fuzz testing identified 7 crashes and 4 hangs, AFL++ identified 10 crashes and 7 hangs, and both methods identified 4 common hangs, for a total of 17 unique crashes and 7 unique hangs. In general, while AFL++ was more effective than classic fuzz testing, classic testing still identified failures not found by AFL++, showing the value of using multiple testing approaches. The results are summarized in Table 2.

### 6.1 cat

The fuzz testing of `cat` did not produce any failures, either in the AI-generated or repository versions. This result is likely due to the program's small size and simple functionality.

### 6.2 dash

The repository version of `dash` had two crashes and no hangs, while the AI-generated version had no failures. The first crash was caused by a global-buffer overflow in `dash`'s character-classification logic. `dash` treats input bytes as signed characters, so bytes with the high bit set can become negative values. These values are then used as indices into lookup tables. Normally, `dash` applies an offset so that a negative value still maps to a valid position within the table. However, one code path failed to apply this offset, so a negative value could be used directly as the index. This is a classic C coding error dating back decades [MFS90]. This use of a signed `char` caused `dash` read outside the table and caused a segmentation fault. The AI-generated implementation avoids this problem by not using signed-character indexing.

The second crash was caused by a strange case that occurs while processing the "`<<`" shell input construct known as a *here-document*. A here-document redirects a block of subsequent input to a command's standard input. The "`<<`" sequence tells the shell to begin the here-document and is immediately followed by the token that specifies the end of the input.

The failure occurred with an input that contained "`<<$(<<x)`". The shell sees the "`<<`" and starts to look for the terminator sequence. However, it finds the "`$(...)`" construct, which is known as a *command substitution.* This construct tells `dash` to execute the text between the parenthesis as shell commands, taking the output from those commands as the substituted value. The strange sequence above starts here-document processing and then, while trying to find the terminator character, re-triggers the processing again (with "`x`" as the terminator character).

The repository version of `dash` uses a single global variable to track the here-document currently being parsed. When it started processing a second here-document input before finishing the first one, the pointer to the structure that tracks the here-document was overwritten. As a result, `dash` never stored the body of the first here-document, leaving it NULL. `dash` later attempted to process this missing body without checking for NULL, resulting in a segmentation fault. The AI-generated implementation avoids this failure because its parser does not parse command substitutions embedded in here-document delimiters; instead, it reports a syntax error and exits cleanly.

## 6.3 dc

Both the repository and AI-generated versions of `dc` had no crashes and one hang. The hang in both versions had the same cause. `dc` supports arbitrary-precision arithmetic. In our test case, the random input contained sequences of the letters A through F (used in hexadecimal numbers), which `dc` accepts as numeric digits even when operating in base 10. As a result, a short sequence such as “EEEEE” is interpreted as a large numeric value. When this value is used with the exponentiation operator “^”, `dc` attempts to compute the complete result using arbitrary-precision arithmetic. The resulting number can contain hundreds of thousands of digits, requiring repeated multiplication of increasingly large values. The computation therefore continues for an extremely long time and is practically indistinguishable from a hang. Given enough time, the calculation should eventually complete, although it may exhaust the available memory first.

## 6.4 grep

`Grep` had one hang in both the repository and AI-generated versions. In both cases, the failure originated in the GNU regular-expression library used by both implementations rather than in `grep` itself. The failure occurs during pattern compilation, before any input is searched. It can be triggered by patterns containing several consecutive “+” quantifiers. GNU’s regex compiler first represents the pattern as a tree. To compile “+”, meaning one or more repetitions of the preceding expression, a copy of the expression’s tree is made. If another “+” follows, it is applied to the entire tree produced by the previous “+”, so this already enlarged tree is copied again. Each additional “+” can therefore double the number of tree nodes, causing exponential growth in memory use and compilation work.

The hangs were detected by AFL++ with a hang timeout of 10 seconds. Given enough time and memory, `grep` would have finished executing, but because each added “+” doubles memory use as well as work, in practice the process exhausts available memory and is killed long before it completes. For example, a pattern containing 27 consecutive “+” quantifiers can require more than 100 GB of memory. It is interesting to note that the “*” is implemented separately from “+” and does not have this problem.

A separate stack-exhaustion case can also occur with such patterns but both repository and AI implementations of `grep` install signal handlers that detect this condition, report a stack overflow error, and exit cleanly rather than crash.

The behavior of “+” has been known in GNU regex implementations for many years. Earlier work documented resource exhaustion problems in GNU regular-expression processing [DO08]. More recently, glibc Bug 29642 [GNU22] reported that multiple adjacent “+” quantifiers rapidly exhaust memory because the compiler creates an exponentially growing number of tree nodes. GNU grep documentation also explicitly warns that some regular expressions can require exponential time and space and may exhaust memory [GREP26]. The glibc security policy recognizes this quadratic or exponential resource consumption during regex compilation or execution and classifies it as a bug rather than a security vulnerability [GNU26].

## 6.5 less

The repository version of `less` had three crashes, while the AI-generated version had 0 failures. These failures all came from classic fuzz testing as AFL++ could not be used to test `less` because `less` requires a pseudo-terminal (PTY), which was not supported by the AFL++ test harness. The missing AFL++ results are indicated as "–" in Table 2.

The three crashes in the repository version were distinct memory-safety errors. The first was an off-by-one array read (heap-buffer-overflow) in the table used to store screen-row positions. One code path accessed an element one position beyond the end of the array, resulting in an out-of-bounds read and a segmentation fault. The second crash occurred during file name completion. A pointer moved backward through a buffer of completion candidates, but the loop examined the preceding byte before confirming that the pointer had not reached the start of the buffer. This caused a read one byte before the allocated region and resulted in a segmentation fault. The third crash was caused by an invalid free in shell-escape handling. The program assumed that a particular pointer always referred to heap-allocated memory, but one execution path assigned it to a string literal. A later call to free therefore attempted to release memory that had not been dynamically allocated, causing the allocator to abort the program.

The AI-generated version avoided all three errors by using different designs and data structures rather than by adding extra checks. It keeps no table of screen-row positions at all, so there is no array to read past the end of. It numbers its completion candidates instead of stepping through them with a pointer, so there is nothing to step past the start of, and a search that matches nothing is rejected before any stepping begins.

## 6.6 make

The repository version of `make` had one crash and no hangs, while the AI-generated version had no failures. The crash in the repository version was caused by unbounded recursive expansion of a macro. A macro in a `Makefile` is a named variable whose value is substituted when it is referenced. `make` already detects the simplest form of self-reference, such as "`A = $(A)`", because while expanding `A` it can see that the same macro is being expanded again and report an error instead of continuing. However, the crash occurred through an indirect reference. For example, with "`f = $(call f)`", expanding the macro `f` invokes "`$(call f)`", which causes `f` to be expanded again. This process repeats indefinitely. As the recursion continues, the call stack keeps growing until the available stack space is exhausted, causing the repository version to terminate with a segmentation fault. The AI-generated version did not crash because it included a limit on how deeply macro expansion could recurse. Once that limit was reached, it stopped the expansion and reported an error instead of exhausting the stack.

## 6.7 ptx

Both the repository and AI-generated versions of `ptx` had one hang and one crash. The hang in the repository version of ptx occurred when a variable became negative. Later logic assumed that this value could not be negative but never checked for it. This caused a loop condition to remain true even though no progress could be made, resulting in an infinite loop.

The hang in the AI-generated version had a different cause. ptx produces one output line for every word occurrence, together with its surrounding context. The AI-generated implementation constructs these lines inefficiently by repeatedly scanning the surrounding text and assembling the output one character at a time. On ordinary newline-bearing input, this behavior remains linear but has a large constant cost: a 100 MB fuzz-generated input took about 10 minutes, compared with about 15 seconds for the repository version. On newline-free input, the problem becomes more severe because the entire file is treated as a single text region. The amount of work performed for each word then grows with the size of the file, producing quadratic behavior. A 2.6 MB newline-free input took more than 10 minutes. These executions therefore appear as hangs even though the program eventually terminates.

The crash in both the repository and AI-generated versions originated in the GNU regular-expression library, which both implementations rely on for regular-expression processing. As discussed for `grep` in Section 6.4, certain complex regular expressions can cause the pattern compiler to construct an extremely large internal representation of the regex. Processing this large representation also involves deep recursion, with one recursive call made for each node along long chains of nodes. This exhausted the available stack and caused a stack overflow. This is a well-known limitation of many regex libraries [DO08].

## 6.8 strings

Neither the repository nor AI-generated versions of `strings` encounter any failures.

## 6.9 tac

The repository version of `tac` had one hang and one crash, while the AI-generated version had none. Both failures in the repository version originated in the GNU regular-expression library used by `tac`. As with `grep` and `ptx`, certain complex regular expressions can cause the pattern compiler to construct an extremely large internal representation of the regex. In one case, this exhausted the available stack and caused a crash. In the other, it led to prolonged execution and was classified as a hang.
The AI-generated version uses the same GNU regular-expression library, but selects a stricter regular-expression syntax. Under this syntax, malformed patterns containing duplicated or stacked quantifiers are rejected during compilation rather than passed to the matcher. As a result, the regular expressions that caused the repository version to crash or hang never reached the vulnerable matching code. This difference was therefore due to the regular-expression syntax selected by the implementation, rather than a difference in the underlying regex library.

## 6.10 tnftp

The repository version of `tnftp` had a total of seven distinct crashes, two found with classic fuzz testing and five with AFL++. The AI-generated version had one crash.

### 6.10.1 `tnftp` Classic Fuzz Crashes (2)

The first crash was not a defect in `tnftp` itself, but in the version of `libedit` used by `tnftp`. `libedit` provides interactive command-line editing. `libedit` produces the number of arguments and a pointer to the argument list (similar to `argc/argv`). However, if the input contains an unmatched single or double quote, the parser's tokenizer fails before these values are initialized. Although the tokenizer reports this failure through its return value, `libedit` internally does not check that result and operates on the uninitialized values. If the uninitialized token count is one or greater, `libedit` treats the uninitialized argument-list pointer as a valid address and attempts to access the first argument through this invalid pointer, causing a segmentation fault. Note that there is no return value from `libedit` that would allow users to check for this kind of error. We previously reported this error in 2022 [MZH22]. Since both the AI-generated and repository versions rely on libedit for command-line editing, this crash was common among them.

The second crash was also in `libedit`. It occurs where `libedit` converts a character into a multibyte representation. The code assumes that five bytes are always enough to store the encoded character. However, in the tested UTF-8 locale, the conversion requires up to six bytes. There is code in `libedit` to catch this problem, but it only calls `abort`. The crash is locale-dependent and it does not occur under the C locale, where characters are processed as single bytes. The repository version of tnftp sets the locale when the required function is available, as determined at build time by a configuration check. This check was enabled in our build, so the repository version adopted our UTF-8 locale and exposed the defective libedit code path. The AI-generated version avoided this crash because it does not set a locale. The manual page does not specify any character-encoding behavior, so its implementation treats input as byte-oriented and remains in the default C locale. The underlying defect is still present in the linked libedit library, but the condition needed to trigger it is absent. The tradeoff is that the AI-generated version silently discards non-ASCII input at its prompt, whereas the repository version accepts it.

### 6.10.2 `tnftp` Coverage Guided Fuzz Crashes (5)

The third crash occurs while `tnftp` processes the `nmap` command, used to automatically rename files during transfers. The command takes an input-pattern and output-pattern. The input pattern is the original file name and the output pattern is the new file name. The crash occurs when `nmap` tries to separate these two patterns. It searches for a space character marking the end of the input pattern. If the two patterns are instead separated by a tab, this search fails and returns a NULL pointer. The code does not check whether the search was successful and immediately tries to replace the separator with a string terminator so that the two patterns become separate strings. Because the pointer is NULL, this write causes a segmentation fault.

The fourth crash was caused by the same underlying `nmap` parsing defect, but it occurred through `tnftp`'s `proxy` command path. As in the previous crash, the `nmap` code searches only for a space character, returning a NULL if one is not found. In the `proxy` command code path, `tnftp` tries to skip past the expected spaces by reading memory using the invalid pointer, leading to a crash.

The fifth crash was caused by unbounded recursion in `tnftp`'s code for handling macros. `tnftp` allows one macro to invoke another, including itself. However, it does not impose a recursion-depth limit or detect cycles between macros. Each recursion consumes additional stack space, so the process eventually exhausts the available stack and terminates with a segmentation fault.The sixth crash also occurred in the macro code. Consider definition for macro `m`:

```
macdef m
0 0
$i
```

When macro `m` is invoked by "`$m 0`", `tnftp` parses the line into three entries: `$`, `m`, and `0`, followed by a NULL, resulting in an argument count of three and a pointer to the shared global array that contains the argument strings. Since this is a shared global array, `tnftp` reuses it whenever it parses another command line. `tnftp` then parses the first line of the macro body, "`0 0`". Because the same global argument array is reused, the previous contents `$` and `m` are overwritten with `0`, `0`, and `NULL`. The argument list has

therefore become shorter, even though the argument count is still three. The command “`0 0`” is invalid because there is no command named `0`. After executing a valid command, the macro-processing code would reparse the original macro call (“`$m 0`”) and restore the argument information associated with it, however the invalid-command path skips this restoration step. As a result, the argument count remains three even though the shared array now contains only two arguments. `tnftp` then processes the second macro line, “`$i`”. “`$i`” is used in a loop to refer to subsequent arguments passed to the macro. In this example, as there have not yet been any loop iterations, it will refer to the first argument passed (which, as described above, was in the third position in the global argument array). However, because of the error when processing the “`0 0`” line, that third element has a NULL pointer. When `tnftp` tries to dereference that pointer to get the value to substitute “`$i`”, there is a segmentation fault.

The seventh crash also occurred in `tnftp`’s code for handling macros. `tnftp` allows a macro to refer directly to an argument passed to it by using `$` followed by a number. To determine which argument was requested, `tnftp` reads the digits following the `$`. This value is stored in a 32-bit signed integer whose largest possible value is 2,147,483,647. However, the code does not check whether the provided number will exceed this limit. With an input such as `$2222222222`, the value eventually exceeds the representable range, causing the signed integer to overflow and become negative. Unfortunately, while the subsequent code performs a bounds check, it only verifies that the requested argument number does not exceed the number of arguments supplied to the macro; it does not check for negative values. `tnftp` uses the negative value as an index into the argument array, causing an out-of-bounds memory access and a segmentation fault.

| Program | Pointers & Arrays | Error handling | Complex State | Included Code | Unbounded Recursion | Algorithmic Complexity |
|---|---|---|---|---|---|---|
| `cat` | | | | | | |
| `dash` | R | | | | | |
| `dc` | | | AR | | | AR |
| `grep` | | | | AR | | |
| `less` | R | | | | | |
| `make` | | | | | R | |
| `ptx` | | | R | AR | | A |
| `strings` | | | | | | |
| `tac` | | | | R | | |
| `tnftp` | R | R | | AR | R | |

**Table 3: Categorization of Failures**

*“A” indicates a failure of the AI-generated version and “R” for the repository version.*

## 6.11 Test Results Summary

We debugged each failure to find its root cause and then (as we have done in previous fuzz studies) categorized the error causes. These results are reported in Table 3. The types of errors found in this study differed in some way from our previous fuzz study that included Linux utility programs. In this study, no failures were caused by careless or missing handling of return values or errors from subprocesses started by the program. On the other hand, we added categories for unbounded recursion and algorithmic complexity.

## 7 A Simple Demonstration of AI as a Path to Software Sustainability

Software sustainability is a major concern on projects with changing staff, low software staffing levels, and staff with limited software experience. To give an indication of how agentic AI software development can help with sustainability, we performed a simple experiment with the AI-generated version of `ptx`. Using the same workflow and skills as during its original development, we asked an agent to replace the GNU regular-expression implementation with RE2, a regular expression engine designed for predictable linear-time matching and to avoid the resource-exhaustion problems we observed with GNU regex. The agent completed the change in approximately 15 minutes. Most interfaces mapped directly, although some configuration changes were needed to preserve behavior. The agent also identified that RE2 does not support the Emacs-style syntax required by two `ptx` options and escalated this incompatibility to the human operator for resolution.

Although small, this experiment shows how the same agentic workflow can support later maintenance. The agent quickly replaced a significant dependency, adapted the surrounding code, and identified a specification conflict requiring human judgment. This suggests that agents may reduce the effort required to sustain and modify software over time.

## 8 Discussion and Conclusions

There are several lessons to be learned from this study.

1. We showed objectively that in the hands of an experienced programmer, agentic AI systems can efficiently produce high quality software. Though we do note that the AI code was subject to hanging, often because of algorithmic errors and inefficiencies. These inefficiencies occurred even though Claude was explicitly instructed to avoid unnecessarily expensive algorithms and data structures.
2. The only crashes observed in the AI-generated software (for `tnftp`) were caused by the use of existing unsafe libraries. In addition, both the repository and AI versions of the software had hangs and crashes in existing libraries that evaluate regular expressions and provide command line editing.
3. In spite of the wide availability of fuzz testing tools, the existing repository code (both the utilities and common libraries) continues to have a surprising number of failures, and certainly more than the AI-generated code. That argues that it is time for increased integration of AI tools into the current Linux software base. We have little doubt that software on other platforms is in a similar state.
4. Stage 5 of our AI workflow, Security and Correctness Audit, was often the most time consuming. This stage could require several iterations until no more issues with the code were found. And sometimes the fix for one issue would create a new issue. This is an area where more research is needed.
5. The evolution of AI models is happening so quickly that the software that they create is only likely to improve. Notably, the effort that it takes for a human directing the process is likely to decrease.
6. We note that fuzz testing continues to be an effective technique for evaluating software. While modern coverage guided testing was more effective than the original black box generational testing, the original technique still exposed failures that the other did not. In an actual software development project, fuzz testing should be integrated into the agentic workflow.

7. When the agentic workflow is delivered with the code – the specifications, the test suites, the prompts and constraints that produced it – then future efforts to modify and update the code by new project staff will not require knowledge of the code structure and internal functionality, lowering the costs and barriers to software sustainability.

It is clear that software development has undergone a massive change in the past couple of years and will continue to change as AI systems continue to evolve. Using best practices, a programmer in charge of an agentic team can produce code of high quality. That agentic AI workflow can also provide the basis for software sustainability over time, allowing for updating and modification of software without knowledge of the internals or functionality of the code. As the AI systems evolve, the role of the programmer may become less and less technical over time.

## Acknowledgements

This material is based upon activities supported by the National Science Foundation under Interagency Agreement #A2407-049-089-064206.0. Any opinions, findings, and conclusions or recommendations expressed are those of the author(s) and do not necessarily reflect the views of the National Science Foundation. We would like to thank Chris Harrison for his participation in the early discussions leading to this project.

## References

[AFL26] AFL++ Project, "Fuzzing with AFL++", AFL++ Documentation, 2026. `https://aflplus.plus/docs/fuzzing_in_depth/`

[Ant25a] Anthropic, "skill-creator/SKILL.md", *Anthropic Skills GitHub Repository*, 2025. `https://github.com/anthropics/skills/blob/main/skills/skill-creator/SKILL.md`

[Ant25b] Anthropic, "Equipping Agents for the Real World with Agent Skills", *Anthropic Engineering*, October 2025. `https://www.anthropic.com/engineering/equipping-agents-for-the-real-world-with-agent-skills`

[Ant25c] Anthropic, "How We Built Our Multi-Agent Research System", *Anthropic Engineering*, June 2025. `https://www.anthropic.com/engineering/multi-agent-research-system`

[BM08] C. Boogerd and L. Moonen, "Assessing the Value of Coding Standards: An Empirical Study", *24th IEEE International Conference on Software Maintenance (ICSM)*, Beijing, China, September 2008.

[CHL+26] J. Chen, H. Huang, Y. Lyu, J. An, J. Shi, C. Yang, T. Zhang, H. Tian, Y. Li, Z. Li, X. Zhou, X. Hu, and D. Lo, "SecureVibeBench: Benchmarking Secure Vibe Coding of AI agents via Reconstructing Vulnerability-Introducing Scenarios", *64th Annual Meeting of the Association for Computational Linguistics,* vol. 1 (Long Papers), San Diego, July 2026.

[CLSZ24] X. Chen, M. Lin, N. Schaerli, and D. Zhou, "Teaching Large Language Models to Self-Debug", *International Conference on Learning Representations (ICLR)*, Vienna, Austria, May 2024.

[DLT+24] Y. Du, S. Li, A. Torralba, J. B. Tenenbaum, and I. Mordatch, “Improving Factuality and Reasoning in Language Models through Multiagent Debate”, *41st International Conference on Machine Learning (ICML)*, Vienna, Austria, July 2024.

[DO08] W. Drewry and T. Ormandy, “Insecure Context Switching: Inoculating Regular Expressions for Survivability”, *2nd USENIX Workshop on Offensive Technologies (WOOT 08)*, San Jose, CA, July 2008.

[Fag76] M. E. Fagan, “Design and Code Inspections to Reduce Errors in Program Development”, *IBM Systems Journal* **15**, 3, September 1976.

[FME+20] A. Fioraldi, D. Maier, H. Eißfeldt, and M. Heuse. “AFL++: Combining Incremental Steps of Fuzzing Research”, *14th USENIX Workshop on Offensive Technologies (WOOT 20)*, online, August 2020.

[GNU22] GNU C Library, “Bug 29642: regcomp with multiple adjacent plus sign would exhaust memory quickly”, GNU C Library Bugzilla, October 2022.
`https://sourceware.org/bugzilla/show_bug.cgi?id=29642`

[GNU26] GNU C Library Project, “Security Policy”, GNU C Library Source Repository, 2026. `https://github.com/bminor/glibc/blob/master/SECURITY.md`

[GREP26] GNU Project, “GNU Grep 3.12 Manual”, GNU Project, 2026. `https://www.gnu.org/software/grep/manual/grep.html`

[HZC+24] S. Hong, M. Zhuge, J. Chen, X. Zheng, Y. Cheng, C. Zhang, J. Wang, Z. Wang, S. K. S. Yau, Z. Lin, L. Zhou, C. Ran, L. Xiao, C. Wu, and J. Schmidhuber, “MetaGPT: Meta Programming for a Multi-Agent Collaborative Framework”, *12th International Conference on Learning Representations (ICLR)*, Vienna, Austria, May 2024.

[JDW+24] X. Jiang, Y. Dong, L. Wang, Z. Fang, Q. Shang, G. Li, Z. Jin, and W. Jiao, “Self-Planning Code Generation with Large Language Models”, *ACM Transactions on Software Engineering and Methodology* **33**, 7, September 2024.

[JS22] E. Jones and J. Steinhardt, “Capturing Failures of Large Language Models via Human Cognitive Biases”, *36th Conference on Neural Information Processing Systems (NeurIPS)*, New Orleans, LA, November 2022.

[KTF+23] T. Khot, H. Trivedi, M. Finlayson, Y. Fu, K. Richardson, P. Clark, and A. Sabharwal, “Decomposed Prompting: A Modular Approach for Solving Complex Tasks”, *11th International Conference on Learning Representations (ICLR)*, Kigali, Rwanda, May 2023.

[KZC+24] A. Kong, S. Zhao, H. Chen, Q. Li, Y. Qin, R. Sun, X. Zhou, E. Wang, and X. Dong, “Better Zero-Shot Reasoning with Role-Play Prompting”, *2024 Conference of the North American Chapter of the Association for Computational Linguistics: Human Language Technologies*, Mexico City, Mexico, June 2024.

[LLH+24] N. F. Liu, K. Lin, J. Hewitt, A. Paranjape, M. Bevilacqua, F. Petroni, and P. Liang, “Lost in the Middle: How Language Models Use Long Contexts”, *Transactions of the Association for Computational Linguistics* **12**, February 2024. DOI: 10.1162/tacl_a_00638.

[MFS90] B.P. Miller, L. Fredriksen, and B. So, "An Empirical Study of the Reliability of UNIX Utilities", *Communications of the ACM* **33**, 12 (December 1990). Also appears (in German translation) as "Fatale Fehlertractigkeit: Eine Empirische Studie zur Zuverlassigkeit von UNIX-Utilities", *iX*, March 1991.

[MTG+23] A. Madaan, N. Tandon, P. Gupta, S. Hallinan, L. Gao, S. Wiegreffe, U. Alon, N. Dziri, S. Prabhumoye, Y. Yang, S. Gupta, B. P. Majumder, K. Hermann, S. Welleck, A. Yazdanbakhsh, and P. Clark, "Self-Refine: Iterative Refinement with Self-Feedback", *37th Annual Conference on Neural Information Processing Systems (NeurIPS)*, New Orleans, LA, December 2023.

[MZH22] B.P. Miller, M. Zhang and E.R. Heymann, "The Relevance of Classic Fuzz Testing: Have We Solved This One?", *IEEE Transactions on Software Engineering* **48**, 6, June 2022, DOI 10.1109/TSE.2020.3047766.

[OB88] T. J. Ostrand and M. J. Balcer, "The Category-Partition Method for Specifying and Generating Functional Tests", *Communications of the ACM* **31**, 6, June 1988.

[PAT+22] H. Pearce, B. Ahmad, B. Tan, B. Dolan-Gavitt, and R. Karri, "Asleep at the Keyboard? Assessing the Security of GitHub Copilot's Code Contributions", *IEEE Symposium on Security and Privacy (SP)*, San Francisco, CA, May 2022.

[PSK+23] N. Perry, M. Srivastava, D. Kumar, and D. Boneh, "Do Users Write More Insecure Code with AI Assistants?", *ACM SIGSAC Conference on Computer and Communications Security (CCS)*, Copenhagen, Denmark, November 2023.

[QLL+24] C. Qian, W. Liu, H. Liu, N. Chen, Y. Dang, J. Li, C. Yang, W. Chen, Y. Su, X. Cong, J. Xu, D. Li, Z. Liu, and M. Sun, "ChatDev: Communicative Agents for Software Development", *62nd Annual Meeting of the Association for Computational Linguistics (ACL)*, Bangkok, Thailand, August 2024.

[Roy70] W. W. Royce, "Managing the Development of Large Software Systems", *IEEE Western Electronic Show and Convention (WESCON)*, Los Angeles, August 1970.

[Roy25] Abhik Roychoudhury, "Agentic AI for Software: Thoughts from Software Engineering Community", arXiv.org, August 2025. `https://arxiv.org/html/2508.17343v1`

[SCD+23] F. Shi, X. Chen, K. Misra, N. Scales, D. Dohan, E. H. Chi, N. Schärli, and D. Zhou, "Large Language Models Can Be Easily Distracted by Irrelevant Context", *40th International Conference on Machine Learning (ICML)*, Honolulu, HI, July 2023.

[SCM+26] H. V. F. dos Santos, V. Costa, J. E. Montandon, and M. T. Valente, "Decoding the Configuration of AI Coding Agents: Insights from Claude Code Projects", *2026 International Workshop on Agentic Engineering (AGENT '26)*, July 2026. DOI 10.1145/3786167.3788412.

[SEI16] Software Engineering Institute, *SEI CERT C Coding Standard: Rules for Developing Safe, Reliable, and Secure Systems*, 2016 Edition, Carnegie Mellon University, June 2016.

[SJS25] S. Shukla, H. Joshi, and R. Syed, "Security Degradation in Iterative AI Code Generation: A Systematic Analysis of the Paradox", *2025 IEEE International Symposium on Technology and Society (ISTAS),* Santa Clara, CA, September 2025. arXiv:2506.11022.

[TDM+25] C. Tony, N. E. Díaz Ferreyra, M. Mutas, S. Dhif, and R. Scandariato, "Prompting Techniques for Secure Code Generation: A Systematic Investigation", *ACM Transactions on Software Engineering and Methodology* **34**, 8, 2025. DOI: 10.1145/3722108.

[Vin26] J. Vincent, "Superpowers: An Agentic Skills Framework & Software Development Methodology", 2026. `https://github.com/obra/superpowers`

[VSM+25] V. Viswanathan, Y. Sun, S. Ma, X. Kong, M. Cao, G. Neubig, and T. Wu, "Checklists Are Better Than Reward Models for Aligning Language Models", 39th *Annual Conference on Neural Information Processing Systems (NeurIPS)*, San Diego, CA, December 2025.

[WBZ+24] Q. Wu, G. Bansal, J. Zhang, Y. Wu, B. Li, E. Zhu, L. Jiang, X. Zhang, S. Zhang, J. Liu, A. H. Awadallah, R. W. White, D. Burger, and C. Wang, "AutoGen: Enabling Next-Gen LLM Applications via Multi-Agent Conversation Framework", *1st Conference on Language Modeling (COLM)*, Philadelphia, PA, October 2024.

[Woo97] M. Wooldridge, "Agent-Based Software Engineering", *IEE Proceedings – Software Engineering* **144**, 1, February 1997.

[WWQ+25] T. Wei, W. Wen, R. Qiao, X. Sun, and J. Ma, "RocketEval: Efficient Automated LLM Evaluation via Grading Checklist", *International Conference on Learning Representations (ICLR)*, Singapore, April 2025.

[WZW+15] E. Wong, L. Zhang, S. Wang, T. Liu, and L. Tan, "DASE: Document-Assisted Symbolic Execution for Improving Automated Software Testing", *37th IEEE/ACM International Conference on Software Engineering (ICSE)*, Florence, Italy, May 2015.

[XY26] R. Xu and Y. Yan, "Agent Skills for Large Language Models: Architecture, Acquisition, Security, and the Path Forward", *Agent Skills '26 Workshop at ACM Conference on AI and Agentic Systems*, San Jose, CA, May 2026.

[YFL+25] R. Yang, F. Ye, J. Li, S. Yuan, Y. Zhang, Z. Tu, X. Li, and D. Yang, "The Lighthouse of Language: Enhancing LLM Agents via Critique-Guided Improvement", *39th Annual Conference on Neural Information Processing Systems (NeurIPS)*, San Diego, CA, December 2025.

[ZCS+23] L. Zheng, W.-L. Chiang, Y. Sheng, S. Zhuang, Z. Wu, Y. Zhuang, Z. Lin, Z. Li, D. Li, E. P. Xing, H. Zhang, J. E. Gonzalez, and I. Stoica, "Judging LLM-as-a-Judge with MT-Bench and Chatbot Arena", *37th Annual Conference on Neural Information Processing Systems* ***36*** *(NeurIPS)*, New Orleans, LA, December 2023.